%% file: vla_quasars_main.tex
\documentclass[fleqn,usenatbib]{mnras}

\usepackage{newtxtext,newtxmath}

\usepackage[T1]{fontenc}

\DeclareRobustCommand{\VAN}[3]{#2}
\let\VANthebibliography\thebibliography
\def\thebibliography{\DeclareRobustCommand{\VAN}[3]{##3}\VANthebibliography}

\usepackage{threeparttable}
\usepackage{lscape, pdflscape}
\usepackage{graphicx}	
\usepackage{amsmath}	
\usepackage{ulem}

\title[Radio-loud fraction of $z > 6$ quasars]{The Radio-Loud Fraction of Quasars at $z>6$}

\input{author-list}

\date{Accepted XXX. Received YYY; in original form ZZZ}

\pubyear{2023}

\begin{document}
\label{firstpage}
\pagerange{\pageref{firstpage}--\pageref{lastpage}}
\maketitle

\begin{abstract}
Quasars at redshifts $z>6$ are an excellent probe of the formation and evolution of supermassive black holes in the early Universe. The population of radio-luminous quasars is of particular interest, as such quasars could potentially be used to study the neutral intergalactic medium  during cosmic reionisation via \textsc{H\,i}~21\,cm  absorption studies. However, the lack of deep radio observations of $z>6$ quasars leaves the population poorly constrained, and suitable candidates for an \textsc{H\,i}~21\,cm absorption study have yet to be found. In this work, we present Jansky Very Large Array (VLA) 1--2~GHz radio continuum observations of 138 quasars at redshifts $6.0 \leq z<7.6$. We detect the radio continuum emission of the $z = 6.1$ quasar J1034-1425, with a 1.6~GHz flux density of $170\pm36\,\mu$Jy. This quasar is radio-quiet with radio-loudness, $R \equiv f_{5\text{~GHz}}/f_{\nu,\text{4400~\AA}} = 2.4\pm0.5$. In addition, we detect 7 other quasars at $z > 6$, which have previously been characterised in the literature at these frequencies. Using the full sample, we estimate the radio-loud fraction to be $3.8^{+6.2}_{-2.4}\%$, where the uncertainties are 95\% confidence intervals. This is lower than recent estimates of the radio-loud fraction in the literature, but is still marginally consistent with no redshift evolution of the radio-loud fraction. We explore the undetected quasar population by stacking their continuum images at their optical positions and obtain a median stacked flux density of $13.8\pm 3.9~\mu$Jy and luminosity of $\log{L_{5~\mathrm{GHz}}/(\mathrm{W~Hz}^{-1})}=24.2\pm0.1$.
\end{abstract}

\begin{keywords}
(galaxies:) quasars: general - (galaxies:) quasars: supermassive black holes - galaxies: high-redshift - galaxies: nuclei – galaxies: evolution - radio continuum: galaxies
\end{keywords}



\section{Introduction}
\label{sec:intro}
Quasars are among the most luminous objects in the Universe, powered by the accretion of matter onto supermassive black holes (SMBHs). Studying quasars at high redshifts ($z>6$) can provide crucial information on the evolution of SMBHs, their host galaxies, and the surrounding large-scale structures in the early Universe. Approximately 10\% of all quasars up to $z\sim5$ are ``radio-loud'', with strong radio synchrotron emission associated with relativistic jets launched from their active galactic nuclei \citep[AGNs; e.g.,][]{Kellermann1989, Kellermann2016, Ivezic2002}. The energy required to launch the powerful jets is thought to originate from the rotational energy of the SMBH \citep{Blandford1977}, the accretion disk \citep{Blandford1982}, or a combination of both. 

A useful measure of the strength of a quasar's radio emission is given by the ``radio loudness'' $R$, which is usually defined as the ratio of the rest-frame luminosities at 5\,GHz and 4400\,\AA\ \citep{Kellermann1989}. A quasar is typically characterised as radio-loud if it has $R>10$, and as radio-quiet otherwise. Traditionally, it has been thought that AGN activity dominates the continuum emission of radio-loud quasars, while radio-quiet emission may be due to a combination of AGN activity and host-galaxy star formation. However, using deep Low Frequency ARray  observations, \hbox{\cite{Gurkan2019}} find that there is no clear separation between radio-loud and radio-quiet quasars i.e., no bimodality in the distribution of $R$, hence raising doubts about whether $R$ carries any physical meaning.

While the definition of $R$ is somewhat arbitrary, the fraction of radio-loud quasars at any given redshift may depend on several factors, including the SMBH accretion modes \citep{Best2012}, the black hole masses and spins, etc \citep[e.g.,][]{Rees1982, Laor2000}. Any evolution in the radio-loud fraction (RLF) would thus indicate evolution in one or more of these properties. Indeed, there has been much debate in the literature as to whether the radio properties of quasars evolve with redshift, with some studies finding that the RLF is a strong function of redshift and optical luminosity and others finding evidence for an unevolving RLF. For example, \cite{Jiang2007} find that the RLF increases with optical luminosity and decreasing redshift, predicting an RLF of about 2\% at $z\approx6$. This is marginally consistent with the findings of \cite{Wang2007}, which, however, were obtained with a small sample size of only 13 quasars. \cite{Kratzer2015} confirm the results of \cite{Jiang2007}, but point out that the observed differences between the behaviour of the RLF and the mean radio-loudness suggest the presence of selection biases. Other studies do not find such correlations, reporting a constant RLF consistent with $\approx 10$\% up to redshifts of $z\sim6$ \citep{Stern2000, Ivezic2002, Banados2015a, Liu2021, Gloudemans2021}. Studies at the highest redshifts have hitherto been limited by small sample sizes, and have hence been unable to rule out redshift evolution. \citet{Banados2015a} stacked the images from the 1.4~GHz Very Large Array FIRST survey \citep{Becker1995} at the positions of 41 quasars at $5.5 < z < 7.1$, finding no significant detection of the stacked continuum emission. Instead, they set an upper limit of 84\,$\mu$Jy to the mean flux density of the undetected quasars at 1.4\,GHz. Clearly, a deeper radio survey of a large quasar sample is needed to detect radio continuum emission from quasars at $z > 6$, and to examine the RLF at these redshifts.

In addition to investigating the SMBH and galaxy evolution, quasars at $z>6$ are excellent tools for studying the Epoch of Reionization (EoR), during which the neutral hydrogen (\textsc{H\,i}) in the Intergalactic Medium (IGM) was ionised by the UV radiation of the first luminous sources. In particular, quasars could potentially be used to probe the neutral IGM via absorption by the forbidden \textsc{H\,i} 21\,cm hyperfine transition \citep{Carilli2002, Carilli2004}. Such a 21\,cm absorption study would be able to probe small-scale structures in the high-$z$ IGM without suffering from many of the systematics and wide-field effects that affect most all-sky 21\,cm emission experiments \citep[e.g.,][]{gmrt, kolopanis2019paper, trott2020mwa, mertens2020lofar, upperlimits2, Singh2022, keller2023}.

The detection of the cosmological \textsc{H\,i} 21\,cm line in absorption requires a strong compact source (150~MHz flux density $\approx 10-100$~mJy) at $z\gtrsim 6.5$ to achieve the necessary sensitivity with modern radio telescopes \citep[e.g.,][]{Ciardi2013}. To date, there are only four known sources at $z>6$ with flux densities exceeding 10~mJy at $\sim$150\,MHz: J1427+3312 \citep{McGreer2006}, J1429+5447 \citep{Willott2010a, Frey2011}, J2318-3113 \citep{Decarli2018, Ighina2021} and J0309+2717 \citep{Belladitta2020}. Of these quasars, J1429+5447 has the highest redshift at $z\approx6.2$, which is marginal to probe the IGM during the EoR. Alternatively, \cite{Thyagarajan2020} proposed to measure the \textsc{H\,i} 21\,cm absorption statistically by stacking the 1-D power spectra of a large number of fainter sources (flux densities $\approx 1-10$~mJy). The selection of a suitable observation strategy requires knowledge about the flux density distribution of the quasar population at these redshifts.

In this study, we used the L-band receivers of the Karl G. Jansky Very Large Array (VLA) to obtain deep 1--2~GHz continuum images of 138 optically confirmed quasars at $z>6$, to measure their radio properties.
Section~\ref{sec:obs} describes our quasar sample and the VLA observations. Section~\ref{sec:imaging} describes the imaging and post-imaging data processing. Our results are presented in Section~\ref{sec:results}, and a summary of this work is provided in Section~\ref{sec:conclusion}.

Throughout this paper, we assume a standard $\Lambda$CDM cosmology with $H_0=67.66$\,km\,s$^{-1}$\,Mpc$^{-1}$, $\Omega_M=0.3$ and $\Omega_\Lambda=0.7$ as measured by \cite{planck}. Magnitudes are given in the AB system.

\section{Data and Observations}
\label{sec:obs}

\subsection{The Quasar Sample}

\begin{figure}
    \centering
    \includegraphics[width=\linewidth]{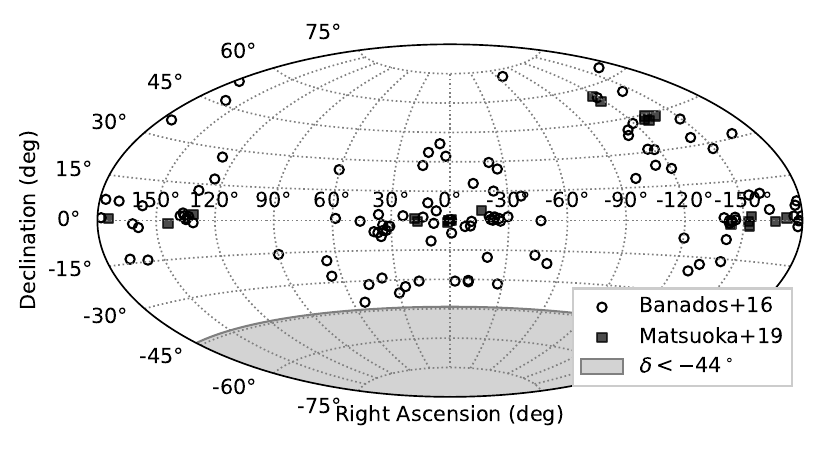}
    \caption{A sky map of the 138 quasars at redshifts $z\geq6$. We restricted the observations to declinations $\delta>-44^\circ$ (shaded regions) for convenience with the VLA.}
    \label{fig:sky_map}
\end{figure}

Our sample of 138 quasars at $z\geq 6$ is based on the sample of the Pan-STARRS1 survey \citep[PS1;][]{Banados2014, Banados2016, Tang2017}, but also includes quasars from a number of other recent optical and near-infrared high-redshift surveys: SDSS \citep[][]{Jiang2009}, CFHQS \citep[][]{Willott2007, Willott2009, Willott2010a, Willott2010b}, UKIDSS \citep[][]{Mortlock2011}, VIKING \citep{Venemans2013}, VST-ATLAS \citep{Carnall2015}, DES \citep{Reed2015, Reed2017}, HSC/SHELLQs \citep{Kashikawa2015, Matsuoka2016, Matsuoka2018, Matsuoka2019b} and DECaLS \citep{Banados2018}. In addition, our sample contains one radio-selected quasar, J1427+3312, which was discovered by \cite{McGreer2006}. Eight quasars from the aforementioned surveys were not included because we restricted the sample to declinations $\geq -44^\circ$, for better observing conditions with the VLA (see Figure~\ref{fig:sky_map}). Except for the declination constraint, the sample included all quasars at $z \geq 6$ that were known at the time of the VLA observations.

\begin{figure}
    \centering
    \includegraphics[width=\linewidth]{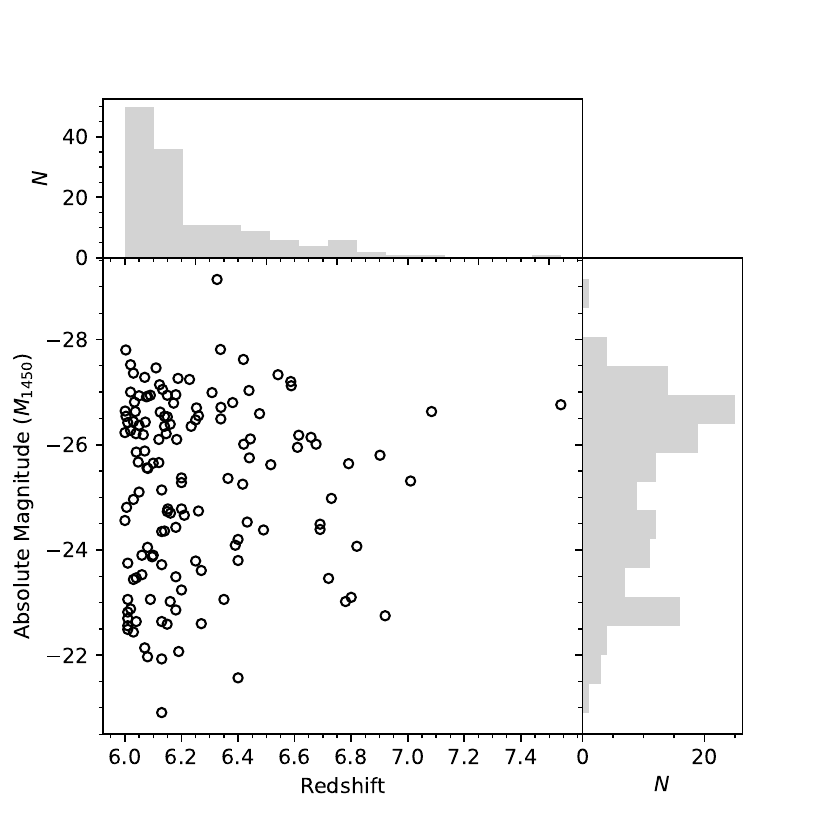}
    \caption{The redshift and absolute rest-frame UV magnitude ($M_{1450}$) distributions of the quasar sample used in this work.}
    \label{fig:M_vs_z}
\end{figure}

The redshift and rest-frame UV magnitude ($M_{1450}$) distributions of the quasar sample are shown in Figure~\ref{fig:M_vs_z}. The quasar redshifts are within the range $6.0\le z < 7.6$, but have a relatively low median of $z=6.15$. The absolute UV magnitudes range from $-29.1$ to $-20.9$, with a median of $-25.6$ and a mode of $\approx -27$. 

\subsection{VLA L-band Observations}
The 1--2~GHz continuum observations were carried out with the VLA between 2019-09-01 and 2019-09-18 (proposal 19A-056, PI: N.~Thyagarajan), using the most extended (A-array) configuration, which has a maximum baseline of 35~km. The A~array was preferred due to its high angular resolution, typically $\lesssim 2^{\prime\prime}$ (depending on source declination and the length of the observation). At $z\sim6$, this corresponds to a projected spatial resolution of about 11~kpc. 

The WIDAR correlator was used as the backend for the observations, with two 512~MHz IF bands covering the frequency range 1--2~GHz. Each IF band was divided into eight 64-MHz sub-bands, each with 64~channels. The data were recorded in full-polar mode, with a time resolution of 2~seconds.

The observing programme was designed to achieve a $5\sigma$ detection threshold of 100~$\mu\mathrm{Jy\,}\mathrm{beam}^{-1}$ at 1.4\,GHz, the approximate depth required to detect prospective candidates for a statistical 21\,cm absorption study (see Section~\ref{sec:intro}). Each quasar target was observed for around 12~min, resulting in a total on-source time of approximately 27.6~h. The observations were grouped in 12 blocks, each starting with 5~min on a primary (flux~density) calibrator and interspersed with regular 2--3~min observations of secondary (phase) calibrators. In total, the programme took up 33~h of observing time at the VLA. In each observing block, flux calibration was performed on 3C48, 3C147, or 3C286, using the flux scale of \cite{Perley2017}.

The data editing and calibration were done with standard tools available from the Common Astronomy Software Applications package \citep[\texttt{CASA} version 6.5.3,][]{casa}. In addition to running the automated \texttt{CASA} tasks, we manually removed time and frequency ranges that are affected by strong radio frequency interference (RFI), resulting in a usable bandwidth of 500--600~MHz at a mean frequency of 1.68~GHz. Lastly, we carefully inspected the various calibration solutions to identify and remove antennas that exhibit noisy gains or other kinds of abnormality. On average, this resulted in about 25 uncorrupted antennas at any given time. 

\section{Imaging and Analysis}
\label{sec:imaging}

The imaging of the VLA L-band data was performed with the $w$-stacking algorithm provided by \texttt{WSClean} \citep[version 3.3,][]{offringa-wsclean-2014, offringa-wsclean-2017}. A subset (about 50\%) of the quasars was imaged independently using the w-projection algorithm \citep{Cornwell2008} in \texttt{CASA}. The results from the two approaches were found to be in good agreement, thus validating our \texttt{WSClean} approach.

For each quasar, we imaged a $\sim 1^\circ$ field of view (FoV) at a pixel resolution of $0.3^{\prime\prime}$, resulting in an image size of $12244\times12244$ pixels. We set the WSClean Briggs robust parameter\footnote{Note that the \texttt{CASA} robust parameter is not exactly the same as the \texttt{WSClean} Briggs parameter \citep{wscleanweighting}.} to $-0.5$, resulting in an average synthesised beam FWHM of $\approx 1.8^{\prime\prime}$. Parallelising over 32 CPUs, the computing time is around 15~min per image. We performed self-calibration on all target fields where an improvement of the signal-to-noise ratio was achievable. Before making the final images, we deployed a set of standard flagging routines on the residual visibilities, but at lower rejection thresholds than in the initial pre-imaging RFI editing. 

Following the imaging, we characterised the statistical properties of the images, which ultimately allow us to determine which images exhibit reliable detections of the quasar radio continuum. The root-mean-square (RMS) of the image noise was robustly estimated as the median absolute deviation (MAD) of the central $41\times 41$~pixel values, multiplied by a scale factor of 1.4826 appropriate for an assumed Gaussian noise distribution \citep{mad}. Using this estimate, we find a median noise RMS of $36\pm 5\,\mu\mathrm{Jy\,}\mathrm{beam}^{-1}$ across the cutout images. 

We fitted a 2-D Gaussian to the central $21\times 21$ pixels of each image, using a least-squares method. We fixed the widths and position angle of the Gaussian to those of the synthesised beam, but let the position vary freely within a range of $\pm3$ pixels. The errors of the fitted amplitudes and positions were determined as described in \cite{gaussian_fitting}. Using the fitted parameters and their associated errors, we used the following two criteria to classify a source as detected: (1) the fitted amplitude is at least three times the noise RMS and (2) the optical position of the target lies within the $2\sigma$ uncertainty contour of the fitted position. In total, there are nine sources that meet these criteria, two of which have not previously been detected at radio frequencies. However, one of these two sources is likely to be a false detection, arising from systematics in the image. This will be discussed in more detail in Section~\ref{sec:results}. Note that there is also a small probability that another source falls within the cross-matching area around the optical position of the target (source confusion). Using the source counts in \cite{source_counts} as a guide, we find a cumulative source density at $S_\nu>100\,\mu$Jy of approximately $10^{-4}$ sources per square arcsec. Hence, our search area of $6\times6$ pixels (3.28~square~arcsecs) contains on average $3\times10^{-4}$ confused sources; source confusion is thus unlikely to significantly affect our results.

Using the assumption that the sources are unresolved, we take the fitted Gaussian amplitudes as a proxy of the total source flux density. Note that the central frequency of the continuum observations is 1.68~GHz. However, without knowledge of the underlying radio spectra, it is not possible to provide an exact frequency corresponding to the measured flux density. Instead, we assume a power-law spectrum ($S_\nu\propto \nu^{\alpha_R}$) for the target quasars, where we adopt a typical spectral index of $\alpha_R=-0.75$. This value is in keeping with previous studies \citep{Wang2007, Momjian2014, Banados2015a, Liu2021} and is in agreement with very long baseline interferometric measurements \citep{Frey2005, Frey2011, Momjian2008, Momjian2018}, which find steep radio spectra for quasars at $z\sim6$. Using this assumed spectrum, we find that our measured continuum flux densities would typically correspond to spectral flux densities at an observed frequency of 1.6~GHz. 

We further transform these flux densities to radio luminosity densities $L_{\nu,\mathrm{5\,GHz}}$ at a rest-frame frequency of 5~GHz by applying a $K$-correction \citep[cf.][]{cosmo_distances, k_correction, Condon2018}

\begin{equation}
\label{eq:lum}
    L_{\nu,\mathrm{5\,GHz}} = 4\pi D_L^2(1+z)^{-(1+\alpha_R)}S_{\mathrm{1.6\,GHz}}\left(\frac{\mathrm{5\,GHz}}{\mathrm{1.6\,GHz}}\right)^{\alpha_R},
\end{equation}
where $D_L$ is the luminosity distance. If a source is detected with $S/N \gtrapprox 10$, it can be imaged in adjacent 64~MHz VLA subbands while retaining sufficient S/N to compute an in-band spectral index (see Section~\ref{sec:spix} in the Appendix for a discussion on calculated in-band spectral indices). For such cases, we perform the extrapolation in equation~\ref{eq:lum} using the fitted power-law rather than the assumed one.

\section{Results}
\label{sec:results}

The flux densities, radio luminosities, and radio-loudness parameters derived from our VLA L-band detections are listed in Table~\ref{tab:results}, along with the redshifts and optical data from the literature. For quasars which are not detected in this study, we provide a table as supplementary material to this paper with $3\sigma$ upper limits on the L-band flux densities and on any other quantities derived therefrom. In total, there are nine quasars detected with $\geq 3\sigma$ significance, 
of which three are radio-loud 
and two do not have any previous radio detections. The two new detections, J0231-2850 ($z=6.0$) and J1034-1425 ($z=6.1$), are found at $3.4\sigma$ and $4.7\sigma$ significance, respectively. However, the radio image of J0231-2850 exhibits clear non-Gaussian artefacts (stripes), and the detection should hence be deemed marginal. To be conservative, we hereafter treat J0231-2850 as a non-detection. The postage stamp cutouts of the L-band continuum images of the eight likely detections and J0231-2850 are shown in Figure~\ref{fig:det_im} of the Appendix. There is no evidence for extended emission in the cutout-images. We checked this more carefully by subtracting the scaled restoring beam from the cutouts and inspecting the residuals around the quasar position. In these residuals, we found no sign of extended emission, thus validating our assumption that the sources are unresolved. 

\subsection{Radio Loudness}
\label{sec:rl}
 The strength of the radio emission of a quasar relative to its optical emission is characterised by the radio-loudness parameter, $R$. There are many definitions of $R$ in the literature \citep[see, e.g., ][for a review]{r_review}. Here, we adopt the most widely used definition $R=f_{5\text{\,GHz}}/f_{\nu,\text{4400~\AA}}$ \citep{Kellermann1989}, where $f_{5\text{~GHz}}$ and $f_{\nu,\text{4400~\AA}}$ are the flux densities at rest-frame 5~GHz and 4400~\AA, respectively. This definition is also in keeping with other high-redshift studies \citep[e.g.,][]{Banados2015a, Liu2021, Gloudemans2021}, therefore allowing a fair comparison with the results therein. A quasar is generally classified as radio-loud if $R>10$, and as radio-quiet otherwise.

At $z=6$, the rest-frame wavelength of 4400~\AA\ corresponds to an observed wavelength of approximately $3\,\mu$m. To mitigate potential large extrapolation errors, $f_{\nu,\text{4400~\AA}}$ should thus be estimated from mid-IR or near-IR observations. In our sample, there are 16 quasars with available photometric data from Spitzer's \textit{Infrared Array Camera} \citep[IRAC,][]{IRAC} at a wavelength of $3.6~\mu$m. There are 49 additional quasars that have $W1$ photometric data in the ALLWISE catalogue of the \textit{Wide-Field Infrared Survey} \citep[WISE,][]{wise} at a wavelength of $3.4~\mu$m. We use these together with a commonly assumed optical spectral index of $\alpha_\nu=-0.5$ \citep[e.g.,][]{Wang2007, Banados2015a} to extrapolate the flux densities to a rest-frame wavelength of 4400~\AA. For the remaining 73 sources, we extrapolate from the AB magnitudes at rest-frame 1450~\AA, using the same spectral index. As pointed out by \cite{Banados2015a}, some quasars might deviate considerably from the average and thus be subject to large extrapolation errors. By comparing the extrapolations from $M_{1450}$ with those from existing IR-photometry, \cite{Banados2015a} estimate the fractional error to be 0.3 for quasars at redshifts $5.5<z<6.5$. We adopted the same error for our flux densities that are extrapolated from $M_{1450}$, with the caveat that there may be individual objects with considerably larger extrapolation errors.
 
\subsubsection{Constraining the Radio-Loud Fraction}

In the following, we use the derived radio-loudness parameters to constrain the fraction of quasars that are radio-loud. In doing this, we replace our radio flux density upper limits with published results from earlier VLA observations, where available. The quasars for which such observations are available are J1148+5251 \citep{Carilli2004}, J0227-0605 \citep{Liu2021} and J1319+0950 \citep{Wang2011}. Furthermore, we include the radio-loud quasar PSO~J172.3556+18.7734 \citep{Banados2021}, which was discovered after our VLA observations were carried out. We exclude the radio-selected source J1427+3312 from the sample, to prevent a bias towards radio-loud quasars.

\begin{figure}
    \centering
    \includegraphics[width=\linewidth]{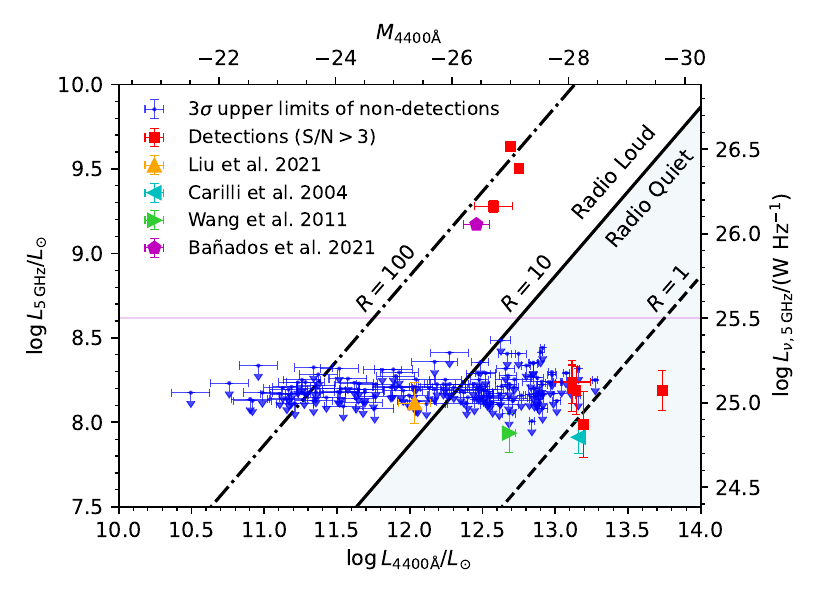}
    \caption{The rest-frame radio luminosity vs. rest-frame optical luminosity at 5~GHz and 4400~\AA\, respectively, of the $z>6$ quasar sample. The blue circles with downward arrows represent $3\sigma$ upper limits and the red squares represent quasars detected in the L-band. The diagonal lines indicate where the radio-loudness ($R$) parameter takes values of 1, 10, and 100 respectively. Radio detected quasars at $R>10$ are classified as radio-loud and all quasars (including non-detections) at $R<10$ are classified as radio-quiet (shaded region). The horizontal magenta line at $\log{L_{\nu,\text{5\,GHz}}}=25.5$~W~Hz$^{-1}$ represents an alternative separation of radio-loud and radio-quiet quasars explored in \protect\cite{Jiang2007} and \protect\cite{Banados2015a}.}
    \label{fig:R}
\end{figure}

Figure~\ref{fig:R} shows the distribution of the derived optical and radio luminosities along with the regions that define radio-loud and radio-quiet quasars, respectively. Quasars with $R<10$ are classified as radio-quiet regardless of whether or not they are detected in our L-band observations. Quasars with $R>10$, on the other hand, can only be conclusively classified as radio-loud if they are detected, since an upper limit in that region cannot rule out the possibility that the quasar is radio-quiet. In total, there are 65 sources that cannot be characterised as radio-loud or radio-quiet. For the remaining sources, 3 are radio-loud detections and 69 are robustly characterised as radio-quiet. Including literature values and excluding the radio-selected quasar J1427+3312, we have in total 65 ambiguous, 4 radio-loud, and 69 radio-quiet quasars. 

Using only sources that can be robustly classified as either radio-loud or radio-quiet, we find a radio-loud fraction of $4/(69+4)=5.5\pm2.7\%$, where the error is estimated from Poisson statistics. If we include the ambiguous quasars and assume that they are all radio-quiet, we obtain a lower estimate of the RLF, $4/138=2.9\pm1.4\%$. This should be considered a lower bound to the value of RLF, as there may be some radio-loud quasars among the ambiguous cases. 

A more rigorous approach to estimating the RLF makes use of survival analysis, which is normally used to analyse the expected time $\tau$ that it takes for a specific event to occur (e.g., lifespan analyses in biology). The probability of an event $E$ not having occurred at time $t$ is given by the Survival function $S(t)=\mathrm{Pr}(\tau>t)$. For our purposes, we replace ``time'' with ``decreasing radio-loudness'', and ``event'' with ``detection''. In other words, we are interested in the cumulative density function $F(r)=\mathrm{Pr}(R<r)=1-S(r)$ of the radio-loudness parameter $r$. We estimate $F(r)$ by employing the Kaplan-Meier estimator \citep[KM,][]{km}

\begin{equation}
    F(r) = \prod_{i: R_i > r}{\left(1-\frac{1}{n_i}\right)},
\end{equation}
where $R_i$ is the radio-loudness parameter of a detected source and $n_i$ is the number of sources that have not been detected at $R > R_i$. This can be shown to be the maximum-likelihood estimator of $F(r)$ \citep[e.g.,][]{Andersen1993}. The advantage of using the KM estimator is that it can deal with censored data, i.e. data where there is only partial information about the exact time (radio-loudness) at which an event (detection) occurs. In this study, we are dealing with left-censored data, or upper limits, which is why we estimate $F(r)$ rather than $S(r)$. The RLF is then simply $\mathrm{Pr}(R>10)=1-F(r=10)$. 

\begin{figure}
    \centering
    \includegraphics[width=\linewidth]{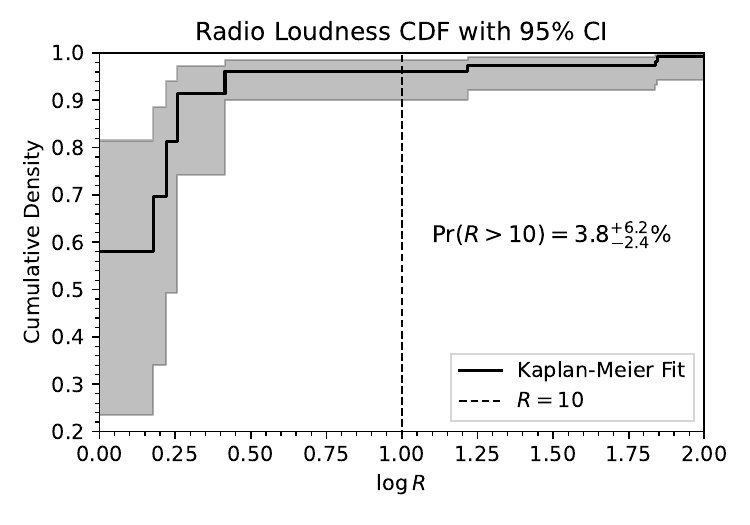}
    \caption{The cummulative density function $F(R)$ of the radio-loudness parameter $R$ as fit by the Kaplan-Meier estimator. The shaded region represents a 95\% confidence interval and the vertical dashed line indicates the radio-loud threshold at $R=10$. The radio-loud fraction is estimated as $\mathrm{Pr}(R>10)=1-F(R=10)=3.8^{+6.2}_{-2.4}\%$.}
    \label{fig:km}
\end{figure}

Our estimate of $F(r)$ is shown in Figure~\ref{fig:km} and was calculated using the \texttt{Python} package \texttt{lifelines} \citep{lifelines}. The RLF thus obtained is $3.8^{+6.2}_{-2.4}\%$, where the uncertainties are 95\% confidence intervals.

\subsubsection{Does the Radio-Loud Fraction Evolve with Redshift?}

\citet{Jiang2007} used a sample of $\gtrsim 30,000$ optically-selected quasars at redsfhits $0<z\leq 5$ to find that the RLF is a strong function of both luminosity and redshift: at a fixed 2500\AA\ absolute magnitude of $-27.7$, they found the RLF to decrease from $\approx 40$\% at $z = 0.5$ to $\approx 8$\% at $z \approx 3$. Conversely, at any redshift, they found the RLF to be higher for more luminous quasars. We note that the median absolute magnitude of our sample at 2500\AA\ is $-26.4$, for an assumed optical-NUV spectral index of $\alpha=-0.5$. However, this median magnitude is close to the lower magnitude limit at which our quasars are robustly classified as radio-loud or radio-quiet. For objects with robust classifications, the effective absolute magnitude of our sample is thus slightly brighter, $\approx -27.7$ at 2500\AA.

\begin{figure}
    \centering
    \includegraphics[width=\linewidth]{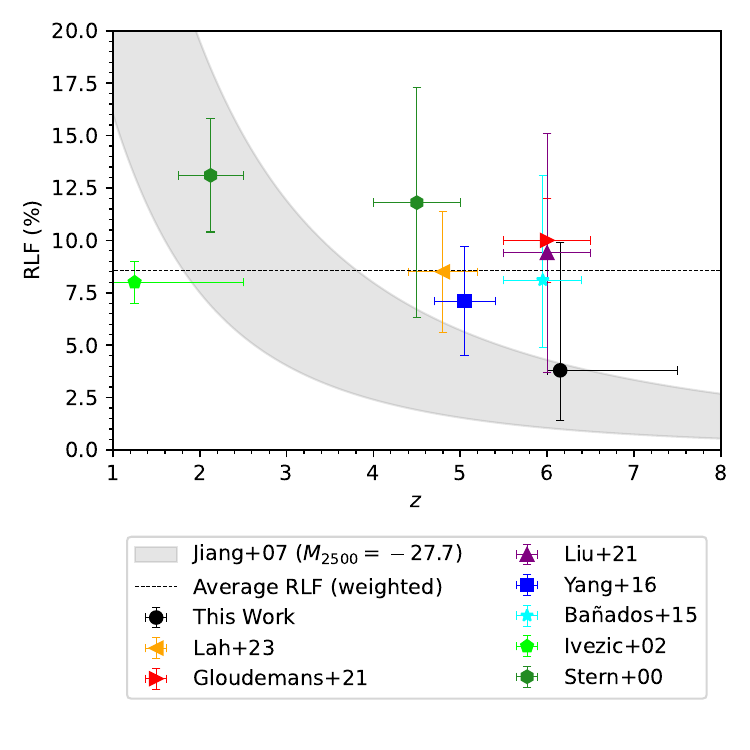}
    \caption{Different estimates of the RLF plotted against redshift. The horizontal error bars indicate the range of redshifts covered by each measurement. The vertical error bars indicate the errors reported in the corresponding studies. Note that uncertainties from the literature are assumed to be $1\sigma$ while ours represents a $95\%$-confidence interval. The marker showing our estimate of the RLF is located at the median redshift of the sample, while the others are placed at the centres of the corresponding redshift ranges. Also shown is is a prediction using the fit from \protect\cite{Jiang2007} at 2500\AA\ absolute magnitudes of $M_{2500}=-27.7$ (grey shaded) and a dashed line representing the inverse variance weighted mean of the RLFs across redshift. See text for discussion.
    \label{fig:rlf_z}}
\end{figure}

Figure~\ref{fig:rlf_z} shows a comparison between our RLF and various literature estimates of the RLF at $z > 4$ \citep{Banados2015a,Yang2016,Liu2021,Gloudemans2021,Lah2023,Ivezic2002,Stern2000}, along with the $1\sigma$ RLF bands predicted by the fit of \citet{Jiang2007} for $M_{2500} = -27.7$. It is clear that our estimate of the RLF is somewhat lower than, but formally consistent with, the earlier estimates of the RLF in the literature. Similar to the earlier studies, we find no significant evidence for redshift evolution in the RLF between $z \approx 4$ and $z \approx 6.2$. However, our measurement of the RLF is also consistent with the prediction of \citet{Jiang2007} of a decline in the RLF with increasing redshift. Deeper observations of a larger sample of quasars at $z \gtrsim 6$ are critically needed to test the above predictions.

\begin{landscape}
\begin{table}
\renewcommand{\arraystretch}{1.1} 
\caption{Radio and Optical Data of the $z>6$ Quasar detections at 1--2\,GHz}
\begin{threeparttable}
\begin{tabular}{lccccccccccc}  
\\\hline\hline
Quasar & RA & Dec. & $z$ & $S_{1.6\mathrm{\,GHz,peak}}$ $^\mathrm{a}$ & $M_{1450}$ & $W_1$ $^\mathrm{b}$ & $S_{3.6\,\mu\mathrm{m}}$ $^\mathrm{c}$ & $\log{L_{\nu, 5\,\mathrm{GHz}}}$ & $\log{L_{4400\,\text{\r{A}}}}$ $^\mathrm{d}$ & $R$ $^\mathrm{e}$ & Ref. $^\mathrm{g}$ \\
& (J2000) & (J2000) & & ($\mu$Jy) & (mag) & (mag) & (mag) & (W~Hz$^{-1}$) & ($L_\odot$) & & (disc./$z$/$M_{1450}$) \\
\hline
J0100+2802 & 01:00:13.02 & 28:02:25.92 & $6.33$ & $101\pm28$ & $-29.1$ & 17.16 $\pm$ 0.03 & - & $25.07\pm0.12$ & $13.74\pm0.01$ & $0.4\pm0.1$ & 1/2/3\\
J0231-2850$^\mathrm{f}$ & 02:31:52.96 & -28:50:20.08 & $6.00$ & $<131$ & $-26.2$ & 20.19 $\pm$ 0.15 & - & $<25.14$ & $12.48\pm0.06$ & $<8.1$ & 4/2/2\\
J0818+1722 & 08:18:27.40 & 17:22:52.01 & $6.02$ & $131\pm32$ & $-27.5$ & - & 18.35 $\pm$ 0.01 & $24.87\pm0.20$ & $13.19\pm0.00$ & $0.8\pm0.4$ & 4/5/3\\
J1034-1425 & 10:34:46.50 & -14:25:15.58 & $6.07$ & $170\pm36$ & $-27.3$ & 18.55 $\pm$ 0.05 & - & $25.07\pm0.14$ & $13.14\pm0.02$ & $1.5\pm0.5$ & 6/6/6\\
J1427+3312 & 14:27:38.59 & 33:12:42.00 & $6.12$ & $1281\pm35$ & $-26.1$ & 19.52 $\pm$ 0.08 & 19.49 $\pm$ 0.02 & $26.39\pm0.01$ & $12.75\pm0.01$ & $76.9\pm2.5$ & 7,8/7/3\\
J1429+5447 & 14:29:52.17 & 54:47:17.70 & $6.18$ & $3003\pm39$ & $-26.1$ & 19.73 $\pm$ 0.08 & - & $26.52\pm0.01$ & $12.69\pm0.03$ & $118.8\pm8.9$ & 9/10/3\\
J1558-0724 & 15:58:50.99 & -07:24:09.59 & $6.11$ & $122\pm36$ & $-27.5$ & - & - & $25.12\pm0.13$ & $13.12\pm0.13$ & $1.8\pm0.8$ & 3/3/3\\
J1602+4228 & 16:02:53.98 & 42:28:24.94 & $6.09$ & $158\pm36$ & $-26.9$ & 18.75 $\pm$ 0.04 & 18.57 $\pm$ 0.02 & $25.09\pm0.14$ & $13.11\pm0.01$ & $1.7\pm0.5$ & 11/5/3\\
J2318-3113 & 23:18:18.35 & -31:13:46.35 & $6.44$ & $637\pm51$ & $-26.1$ & - & - & $26.16\pm0.03$ & $12.58\pm0.13$ & $68.8\pm21.3$ & 12/12/12\\
\hline
\end{tabular}
\begin{tablenotes}
    \item \textbf{Notes.} The full table including all the upper limits are provided as supplementary material to this paper.
    \item $^\mathrm{a}$ L-band continuum observations are typically reported at 1.4\,GHz. However, the exact frequency associated with the flux density in the synthesised image cannot be known without knowledge of the spectral properties of the sources in the map. The central frequency of our observations is 1.68\,GHz, but for a typical source with spectral index $\alpha_R=-0.75$, we find 1.6\,GHz to be a more representative frequency of the reported flux densities (cf. Section~\ref{sec:imaging}).
    \item $^\mathrm{b}$ The $W_1$ magnitudes are taken from the ALLWISE catalogue when available.
    \item $^\mathrm{c}$ The $S_{3.6\,\mu\mathrm{m}}$ flux densities are taken from \cite{Leipski2014}.
    \item $^\mathrm{d}$ The luminosity at rest frame $4400\,\text{\r{A}}$ is obtained by extrapolating $S_{3.6\,\mu\mathrm{m}}$, $W_1$ or $M_{1450}$ in that order of availability (see Section~\ref{sec:rl}).
    \item $^\mathrm{e}$ We use the definition $R=f_{5\text{\,GHz}}/f_{\nu,\text{4400\,\AA}}$ of radio loudness \citep[see Section~\ref{sec:rl} and][]{Kellermann1989}.
    \item $^\mathrm{f}$ The detection of J0231-2850 is not reliable, as the image shows non-thermal-like artefacts. We therefore treat it as a non-detection and leave its confirmation to future work. The upper limit provided for this source is at $3\sigma$ significance.
    \item $^\mathrm{g}$ The references to the discovery papers, and the papers reporting the values of $z$ and $M_{1450}$.
    \item \textbf{References. }(1) \cite{Wu2015}, (2) \cite{Wang2016}, (3) \cite{Banados2016}, (4) \cite{Fan2006a}, (5) \cite{Carilli2010}, (6) \cite{Chehade2018}, (7) \cite{McGreer2006}, (8) \cite{Stern2007}, (9) \cite{Willott2010a}, (10) \cite{Wang2011}, (11) \cite{Fan2004}, (12) \cite{Decarli2018}.
    
\end{tablenotes}
\end{threeparttable}
\label{tab:results}
\end{table}
\end{landscape}

\subsection{Constraining the Undetected Source Population}

\begin{figure*}
    \centering
    \includegraphics[width=0.95\linewidth]{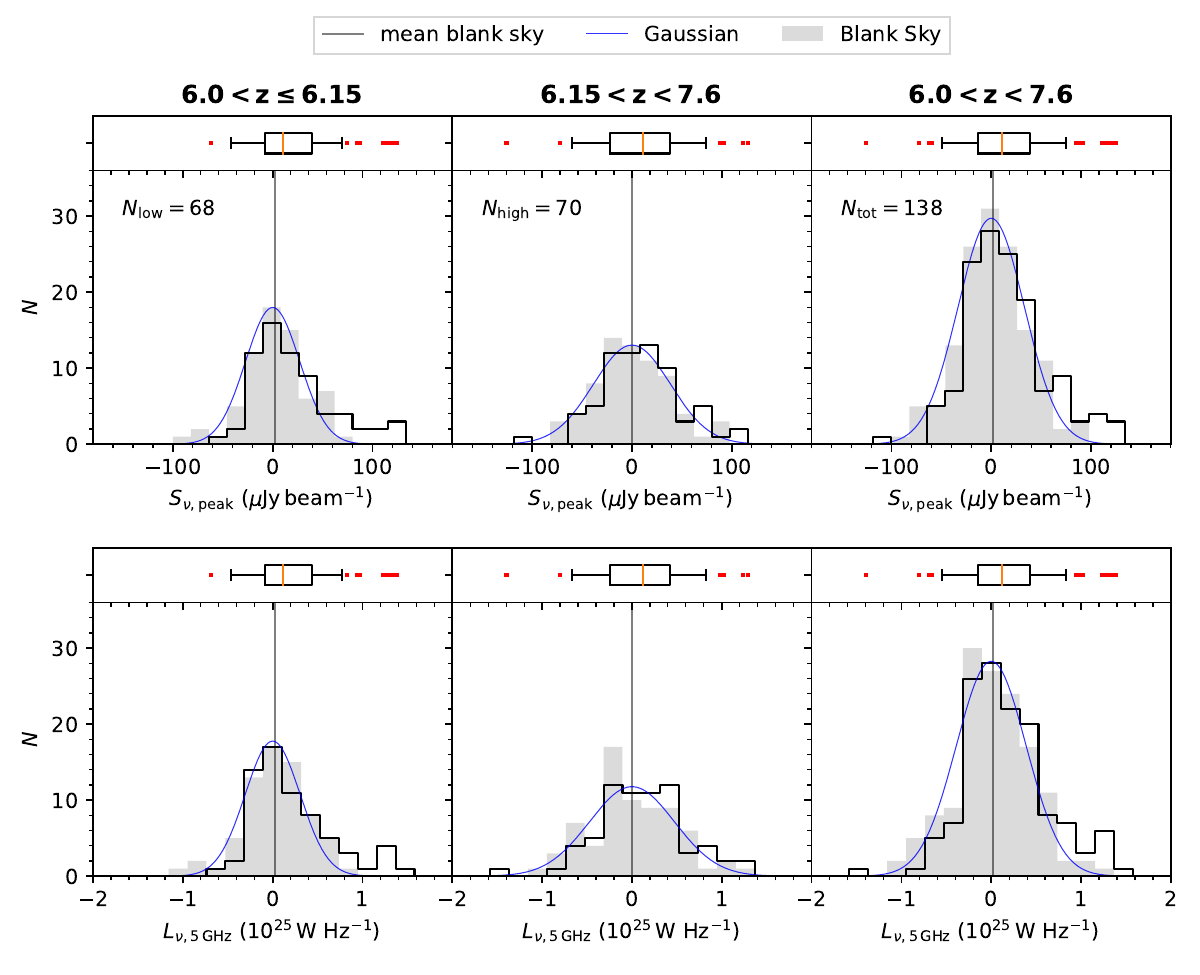}
    \includegraphics[width=0.95\linewidth]{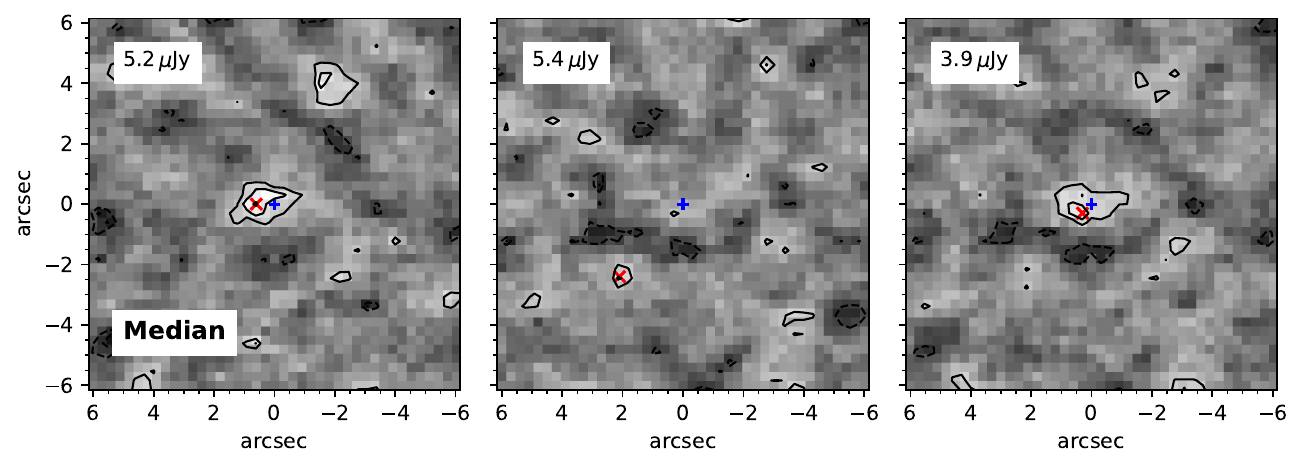}
    \caption{The distributions of the flux density $S_{\nu,\text{peak}}$ (top) and 5~GHz rest frame luminosity $L_{\nu,\mathrm{5~GHz}}$ (middle row) computed from our 1--2~GHz continuum images at the optical positions of the $z>6$ quasars. The three columns correspond to the redshift ranges $6.0<z\leq6.15$ (left), $6.15<z<7.6$ (centre), and $6.0<z<7.6$ (right), respectively. The grey-shaded region shows the amplitude distribution obtained on the ``blank-sky'' regions located $1^{\prime}$ from the optical positions of the quasars. The grey solid vertical lines indicate the means and the blue solid lines the best-fit Gaussians of the blank-sky distributions. The inserted box whisker plots show the medians (orange vertical line), the interquartile ranges (IQR, box), the $1.5\times\mathrm{IQR}$ (whiskers) and outliers (red markers). The bottom row shows the median images of the corresponding redshift ranges. The contour levels correspond to $2\sigma$ and $3\sigma$ with negative values shown as dashed lines. Note that the detected sources are outside the range of these plots but still contribute to the summary statistics. The fitted values and uncertainties of the stacked quantities are listed in Table~\protect\ref{tab:stacking}.}
    \label{fig:stacking}
\end{figure*}

In the context of radio surveys, ``stacking'' is used to glean information on sources that lie below the flux density detection threshold of the survey, but which have been identified at other wavelengths (e.g., infrared, optical, X-ray, etc). This enables the investigation of the statistical properties of the source population below the survey detection threshold. Stacking sources in bins of different parameters (e.g. redshift, stellar mass, colour, etc), allows us to investigate how the average source properties depend on these parameters. Further, the average flux density of individually undetected sources serves as a useful guide for deeper future surveys. In this section, we investigate the evolution of the average quasar radio properties with redshift.

Several methods of stacking radio continuum emission have been proposed in the literature. \cite{White2007} compared the stack mean with the stack median, concluding that the latter is preferable due to its lower sensitivity to the high flux density tail of the distribution, as well as higher robustness towards potential systematic effects (e.g. confusion). They further show that the median approaches the mean in the limit of a noise-dominated distribution. Other authors \citep[e.g.,][]{Mahony2012, Condon2013, Zwart2015} point out several limitations of stacking such as the bias towards flux densities just below the detection threshold. As an alternative, they suggest fitting parametric models to the flux density distribution without the use of a single summary statistic. This ``stack-fitting'' has been further developed and employed in \cite{Mitchell-Wynne2014}, \cite{Rosebook2014}, and \cite{Malefahlo2020}.

As we are dealing with only 138 quasars, which is small relative to the studies mentioned above, we did not attempt to fit the flux density distribution and leave this as a possibility for future work. Instead, we stack the flux density at the positions of the optical quasars, using median-stacking. As a reference, we also stack the flux densities at ``blank-sky'' positions located $\sim$$1^{\prime}$ from the optical positions. We stack in two separate redshift bins, $6.0<z\leq6.15$ and $6.15<z<7.6$ (low-$z$ and high-$z$, hereafter, respectively), which were chosen so that both bins contain a similar number of sources ($N_\mathrm{low}=68$ and $N_\mathrm{high}=70$). The median redshifts of the bins are 6.06 and 6.37, respectively, corresponding to a time difference of approximately 60~Myr. 

For each quasar, we measured the flux density and the luminosity at both the optical quasar position and the ``blank-sky'' position. The flux density and luminosity distributions of the quasars in the two redshift bins, and those of the complete sample (all-$z$, hereafter), are shown as the black histograms in the upper two panels of Figure~\ref{fig:stacking}. The blank-sky distributions are shown as the grey shaded regions; the vertical grey solid line in each panel indicates the mean of the corresponding blank-sky distribution, while the the blue solid curve shows the Gaussian fit to the blank-sky distribution. Both the low-$z$ and the all-$z$ bins show an apparent excess of flux density at the quasar positions, compared to the blank-sky distribution. This suggests that there may be several sources at $z \approx 6$ just below the detection threshold of this survey. The apparent excess is not seen in the high-$z$ sample. In passing, we note that a two-sided Kolmogorov-Smirnov test finds that the null hypothesis that the flux densities at both the quasar positions and the blank-sky positions are drawn from the same distribution cannot be rejected at high significance (p-values of 0.17, 0.5, and 0.15 for the low-$z$, high-$z$, and all-$z$ samples, respectively).

The lowest panel of Figure~\ref{fig:stacking} show the median-stacked images of the low-$z$, high-$z$, and all-$z$ samples. The stacked images for the low-$z$ and all-$z$ bins have RMS noise values of $\approx 5.2\, \mu$Jy and $\approx 3.9\, \mu$Jy, respectively, and show detections of the stacked continuum signal at $\mathrm{S/N}>3\sigma$ significance. For the high-$z$ bin, the stacked image has an RMS noise of $\approx 5.4\, \mu$Jy, and does not show any evidence for statistically-significant emission at the image centre.

\begin{table}
    \renewcommand{\arraystretch}{1.1} 
    \centering
    \caption{Stacking results for the $z>6$ quasar sample. The columns correspond to the three redshift bins used for stacking and are defined by the minimum, maximum and median redshifts $z_\mathrm{min}$, $z_\mathrm{max}$, $z_\mathrm{med}$, respectively. The RMS noise is that of the stacked image cutouts, $S_{\nu\mathrm{,peak}}$ is the fitted peak flux density of the median stacked images, $L_{5~\mathrm{GHz}}$ is the fitted median radio luminosity at rest-frame 5~GHz, and $R$ is the fitted median radio-loudness parameters. The uncertainties of the stacked parameters correspond to the RMS noise in the stacked images. Upper limits represent $3\sigma$ uncertainty levels.}
    \begin{threeparttable}
    \begin{tabular}{lccc}
        \\\hline\hline
            $(z_\mathrm{min}, z_\mathrm{max})$ & (6.0, 6.15) & (6.15, 7.6) & (6.0, 7.6) \\
            $z_\mathrm{med}$ & 6.06 & 6.39 & 6.15 \\\hline
            \multicolumn{4}{l} {\textbf{Median}} \\
            RMS ($\mu$Jy\,beam$^{-1}$) & 5.2 & 5.4 & 3.9  \\
            $S_{\nu\mathrm{,peak}}$ ($\mu$Jy~beam$^{-1}$) & $20.6 \pm 5.2$ & $<16.2$ & $13.8 \pm 3.9$  \\
            $\log{L_{5~\mathrm{GHz}}}/($W~Hz$^{-1}$) & $24.4\pm0.1$ & $<24.3$ & $24.2 \pm 0.1$  \\
            $R$ & $1.1\pm0.3$ & $<0.84$ & $0.7\pm0.2$  \\\hline
            \multicolumn{4}{l} {\textbf{Blank Sky Mean}} \\
            $S_{\nu\mathrm{,peak}}$ ($\mu$Jy\,beam$^{-1}$) & $2.6 \pm 3.9$ & $0.4 \pm 4.5$ & $1.5 \pm 3.0$  \\
        \hline\hline
    \end{tabular}
    \end{threeparttable}
    \label{tab:stacking}
\end{table}

Table~\ref{tab:stacking} lists the flux densities, the rest-frame 5-GHz radio luminosities, and the radio loudness parameters obtained by fitting a single Gaussian to each median-stacked image. We also list the corresponding mean values for the blank-sky stacked images. The median stacked L-band flux density is $S_{\nu\mathrm{,peak}}=13.8\pm3.9$~$\mu$Jy, while the median rest-frame 5~GHz radio luminosity of the quasar sample is $\log{L_{5~\mathrm{GHz}}/(\mathrm{W~Hz}^{-1})}=24.2\pm0.1$. We obtain a median radio-loudness parameter of $R=0.7\pm0.2$ for the entire sample, confirming that the undetected quasar population at $z \gtrsim 6$ is predominantly radio-quiet. As expected, the mean blank-sky flux densities are consistent with zero, underlining that the observed flux density excess at the quasar positions is likely to indeed be due to an undetected population of radio sources.

\section{Summary}
\label{sec:conclusion}

We report VLA 1--2~GHz continuum imaging of a sample of 138 quasars at $z \approx 6.0-7.6$, aiming to determine their radio properties, search for possible redshift evolution in the radio-loud fraction, and identify good targets for follow-up searches for redshifted \textsc{H\,i}~21\,cm absorption. Our search revealed a new radio-quiet quasar, J1034-1425, with flux density $170\pm36$\,$\mu$Jy at redshift $z=6.1$, and detected seven additional quasars that have previously been characterised at radio wavebands. For those sources, our results agree well with literature measurements and are consistent with no temporal variability. We find no new radio sources with 1.6~GHz flux densities $\gtrsim 1$~mJy amongst the sample of 138 quasars, i.e. no immediately suitable targets for follow-up \textsc{H\,i}~21\,cm absorption spectroscopy. 

Our survey robustly classifies 3 quasars as radio-loud (R~$> 10$) and 69 as radio-quiet, while the remaining 66~quasars are indeterminate as our upper limits on their L-band flux density do not rule out their being radio-loud systems. Using the Kaplan-Meier estimator, we find a radio-loud fraction of $3.8^{+6.2}_{-2.4}\%$, where the uncertainties represent $95\%$ confidence intervals. Due to our large sample size relative to that of \cite{Banados2015a} and \cite{Liu2021}, this is the most reliable measurement of the RLF at $z>6$ to date. However, while the central value of our RLF is lower than earlier estimates at $z \gtrsim 4$, our measurement is formally consistent with the earlier studies within $\approx 2\sigma$ significance. Our results are thus consistent with an unchanging RLF at $z > 4$. 

By stacking our continuum images, we find that the undetected quasar population has a median flux density of $13.8\pm 3.9~\mu$Jy, a factor of $\approx 3$ below our typical RMS noise values. Significantly deeper observations would be needed to detect the L-band continuum emission from the bulk of the quasar population at these redshifts. Studies at these redshifts are hence likely to continue to be limited to the brightest quasars in the foreseeable future. However, due to the ongoing optical search for high-redshift quasars, the number of quasar identifications at $z>6$ is steadily increasing \citep[e.g., see new discoveries by DESI;][]{Yang2023}{}{}. According to the Million Quasar Catalogue \citep[MILLIQUAS,][]{milliquas2023}{}{}, there are now 308 known quasars at $z\geq6$ as of August 2023, more than double the number observed in this work. This will allow future surveys to achieve much tighter constraints on the RLF, which should allow a definitive conclusion on its possible redshift evolution.

\section*{Acknowledgements}
The National Radio Astronomy Observatory is a facility of the National Science Foundation operated under cooperative agreement by Associated Universities, Inc.

We thank the anonymous referee for their feedback and comments, which improved the quality of this paper. We also thank Bojan Nikolic and Chris Carilli for valuable inputs.

The software developed for this analysis uses Python and the publicly-accessible and open-sourced Python packages \texttt{Numpy} \citep{numpy}, \texttt{SciPy} \citep{scipy}, \texttt{Astropy} \citep{astropy}, \texttt{Matplotlib} \citep{matplotlib} and \texttt{lifelines} \citep{lifelines}. We furthermore acknowledge the use of \texttt{CASA} \citep{casa} for data editing and calibration, \texttt{WSClean} \citep{offringa-wsclean-2014, offringa-wsclean-2017} for imaging, and \texttt{AOFlagger} \citep{Offringa2010, Offringa2012} for RFI excision.

P.M. Keller is funded by the Institute of Astronomy and Physics Department of the University of Cambridge via the Isaac Newton Studentship. AK and NK acknowledge the Department of Atomic Energy for funding support, under project 12-R\&D-TFR-5.02-0700. 

\section*{Data Availability}

The full data table of this work (Table~\ref{tab:results}) is available in machine-readable form and is provided as supplementary material to this article. The raw data from the VLA observing programme 19A-056 is publicly available from the NRAO data archive at \url{https://data.nrao.edu/}.



\bibliographystyle{mnras}
\bibliography{bibliography} 




\appendix

\section{In-band Spectral Index Fitting}
\label{sec:spix}

The radio-spectra of astronomical sources are typically parameterised by a power-law $S_\nu= S_{\nu_0}(\nu/\nu_0)^{\alpha_R(\nu)}$, where $\alpha_R(\nu)$ is a frequency-dependent spectral index, and $S_{\nu_0}$ is the flux density at frequency $\nu_0$. When extrapolating or interpolating to another frequency over a small fractional bandwidth, one usually assumes a constant spectral index. 

The problem of extrapolating from a continuum measurement is two-fold. Firstly, the local spectral index is not necessarily known and must be either measured or given an assumed value that is representative of the source population under investigation. Secondly, the frequency corresponding to the source flux density in the synthesised continuum image may not be the same as the central frequency of the image if the spectrum is not flat (i.e., $\nu_0$ is uncertain). For bright sources, these uncertainties can be mitigated by dividing the band into sub-bands, which are imaged separately. The extracted flux densities can then be used to fit for $\nu_0$ and $\alpha_R$.

\begin{figure}
    \centering
    \includegraphics[width=\linewidth]{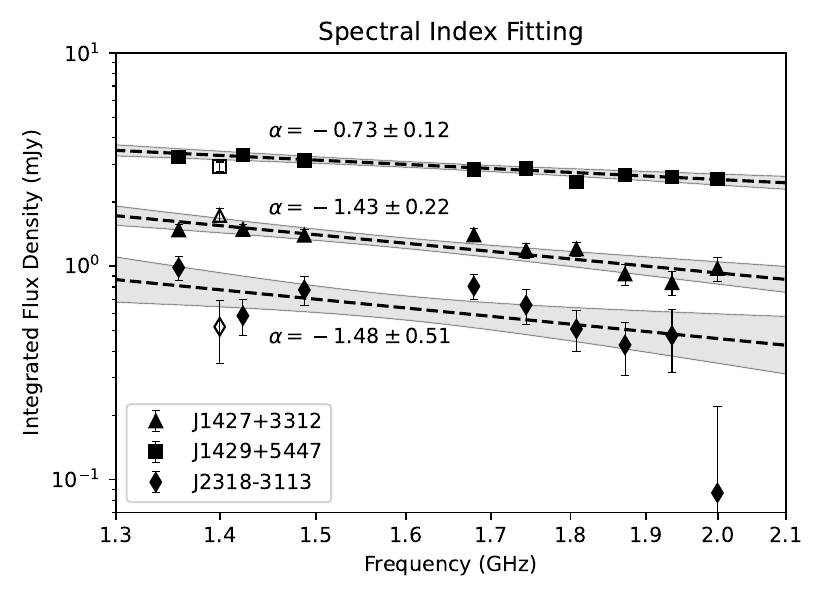}
    \caption{The L-band measurements of the spectra of the quasars J1427+3312 (triangles), J1429+5447 (squares) and J2318-3113 (diamonds). The dashed lines represent the best-fit power laws and the grey shaded regions the corresponding 95\% confidence intervals. The markers with no fill represent literature values (see Table~\ref{tab:spix}.)}
    \label{fig:spix}
\end{figure}

In this work, we have three quasars that are bright enough to fit their in-band spectra: J1427+3312, J1429+5447 and J2318-3113. We image nine spectral windows, each of which has a width of approximately 64~MHz, and is relatively free of RFI contamination. The flux densities within the spectral windows are computed as described in Section~\ref{sec:imaging} and are plotted in Figure~\ref{fig:spix}. For these sources, we use our fitted power-law values to obtain the results listed in Table~\ref{tab:results}.

Table~\ref{tab:spix} lists the fitted values for the flux density and the spectral index at the observed frequency of 1.4~GHz, together with values from the literature (references in the table notes). All three sources have fairly steep spectra. The literature values for J1427+3312 and J1429+5447 are consistent (within $2\sigma$ significance) with our measurements. There are, nevertheless, several effects that might cause the results to differ. Firstly, the spectral indices from the literature were computed from two spectral measurements separated by several GHz in each case (see the table notes in Table~\ref{tab:spix}). Therefore, the results could be biased at 1.4~GHz in cases where $\alpha_{R}(\nu)$ varies across frequency. Secondly, the spectral index of J1429+5447 was determined from VLBI measurements \citep{Frey2011}, which might be resolving out some of the radio emission and would thus be more sensitive to the compact core, which typically has a flatter spectrum. However, the spectral index found in \cite{Frey2011} is in fact steeper than the one obtained in this work, thus ruling out this possibility. Lastly, temporal variability may also be at play with some of the sources. While the literature has not yet noted any variability for J1427+3312 and J1429+5447, there have been inconsistencies for J2318-3113 at lower frequencies (888\,MHz) hinting at variability \citep{Ighina2021, Ighina2022}. Our measurements are consistent with literature values reported at similar frequencies, so we do not find any strong evidence for variability of these sources in the L-band (see Figure~\ref{fig:spix} and Table~\ref{tab:spix}).

\begin{table}
    \renewcommand{\arraystretch}{1.1} 
    \centering
    \caption{Spectral index fitting parameters. For each quasar the first row corresponds to the fitted parameters from this work and the second row to the values found in the literature (see table notes for the references).}
    \begin{threeparttable}
    \begin{tabular}{@{}l@{\hspace{20pt}}c@{\hspace{20pt}}c@{\hspace{20pt}}c@{}}
        \\\hline\hline
        Quasar & $S_{1.4\mathrm{\,GHz,peak}}$ (mJy) & $\alpha_R$ & ref. \\\hline
        J1427+3312 & 1.55$\pm$0.06 & $-1.43\pm0.22$ & 0/0\\
                   & 1.73$\pm$0.13 & $-1.1$ & 1/2$^\mathrm{a}$\\
        J1429+5447 & 3.31$\pm$0.08 & $-0.73\pm0.12$ & 0/0\\
                   & 2.93$\pm$0.14 & $-1.0$ & 1/3$^\mathrm{b}$ \\
        J2318-3113 & 0.78$\pm$0.07 & $-1.48\pm0.51$ & 0/0\\
                   & 0.52$\pm$0.17 & $-1.24\pm0.04$ & 4/5$^\mathrm{c}$ \\
        \hline
    \end{tabular}
    \begin{tablenotes}
        \item \textbf{References}. (0) this work, (1) \cite{FIRST}, (2) \cite{Momjian2008}, (3) \cite{Frey2011}, (4) \cite{McConnell2020} and (5) \cite{Ighina2022}
        \item[a] VLA observations at 1.4\,GHz and 8.4\,GHz.
        \item[b] VLBI observations at 1.5\,GHz and 5\,GHz.
        \item[c] Simultaneous ATCA observations at 2.1\,GHz, 5.5\,GHz and 9\,GHz.
    \end{tablenotes}
    \end{threeparttable}
    \label{tab:spix}
\end{table}

\section{VLA L-Band Images}

\begin{figure*}
    \centering
    \includegraphics[width=0.85\linewidth]{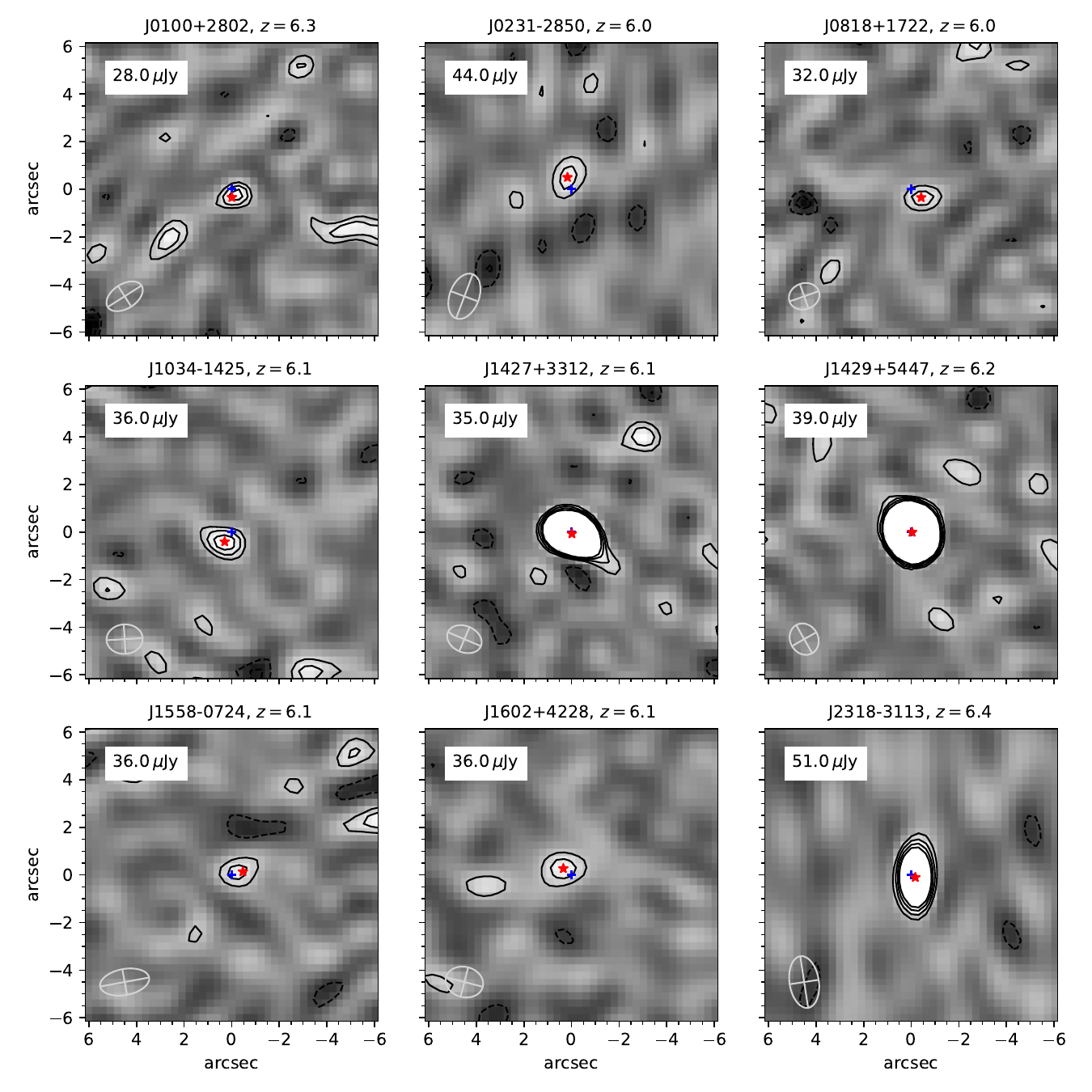}
    \caption{Image cutouts of the nine quasars detected in our 1--2~GHz continuum observations. J0231-2850 exhibits artefacts (lines) and is henceforth treated as a non-detection. The values shown in the upper left corner of the cutouts show robust estimates of the noise RMS (cf. Section~\ref{sec:imaging}). The contour lines are at the level of $2$, $3$, $4$ and $5$ times the noise RMS. Negative contours are shown as dashed lines. The blue cross marks the optical position, and the red star marks the best-fit radio position of the quasar. The ellipse in the bottom-left corners of the cutouts indicates the synthesised beam size (FWHM).}
    \label{fig:det_im}
\end{figure*}

Figure \ref{fig:det_im} shows postage stamp cutouts of the L-band
continuum images of the eight likely detections and J0231-2850, which is treated as a marginal case due to the presence of non-thermal-like artefacts in its image. These detections are discussed in more detail in the results section (Section \ref{sec:results}).

\bsp	
\label{lastpage}
\end{document}

%% file: author-list.tex
\author[P.\ M.\ Keller et al.]{	Pascal M. Keller,$^{1,2}$\thanks{Email: pmk46@cam.ac.uk}
	Nithyanandan Thyagarajan,$^{2}$
    Ajay Kumar,$^3$
    Nissim Kanekar,$^3$
	Gianni Bernardi,$^{4,5,6}$
\newauthor
\\
$^{1}$ Cavendish Astrophysics, University of Cambridge, Cambridge, CB3 0HE, UK\\
$^{2}$ Commonwealth Scientific and Industrial Research Organisation (CSIRO), Space \& Astronomy, P. O. Box 1130, Bentley, WA 6102, Australia\\
$^{3}$ National Centre for Radio Astrophysics, Tata Institute of Fundamental Research, Pune University, Ganeshkhind, Pune 411007, India\\
$^{4}$ INAF-Istituto di Radioastronomia, via Gobetti 101, 40129 Bologna, Italy\\
$^{5}$ Department of Physics and Electronics, Rhodes University, PO Box 94, Grahamstown, 6140, South Africa\\
$^{6}$ South African Radio Astronomy Observatory, Black River Park, 2 Fir Street, Observatory, Cape Town, 7925, South Africa\\
}